# Protein-DNA binding sites prediction based on pre-trained protein language model and contrastive learning

Yufan Liu[1] and Boxue Tian[1*]

[1]MOE Key Laboratory of Bioinformatics, State Key Laboratory of Molecular Oncology, School of Pharmaceutical Sciences, Tsinghua University, Beijing, 100084, China

*Corresponding: boxuetian@tsinghua.edu.cn

## Abstract

Protein-DNA interaction is critical for life activities such as replication, transcription, and splicing. Identifying protein-DNA binding residues is essential for modeling their interaction and downstream studies. However, developing accurate and efficient computational methods for this task remains challenging. Improvements in this area have the potential to drive novel applications in biotechnology and drug design. In this study, we propose a novel approach called CLAPE, which combines a pre-trained protein language model and the contrastive learning method to predict DNA binding residues. We trained the CLAPE-DB model on the protein-DNA binding sites dataset and evaluated the model performance and generalization ability through various experiments. The results showed that the AUC values of the CLAPE-DB model on the two benchmark datasets reached 0.871 and 0.881, respectively, indicating superior performance compared to other existing models. CLAPE-DB showed better generalization ability and was specific to DNA-binding sites. In addition, we trained CLAPE on different protein-ligand binding sites datasets, demonstrating that CLAPE is a general framework for binding sites prediction. To facilitate the scientific community, the benchmark datasets and codes are freely available at https://github.com/YAndrewL/clape.

# Introduction

The interaction of protein and ligands dominate almost all the life activities in organisms, including interactions of protein-protein, protein-small molecules, and protein-nucleic acids. As carriers of genetic information, DNA molecules binding with proteins play a crucial role in many biological processes, including DNA transcription, replication, expression, signal transduction, and metabolism[1,2]. In prokaryote and eukaryote species, approximately 3% and 7% of genomes encode DNA-binding proteins, respectively[3]. Transcription factors (TFs) are a representative group of DNA-binding proteins that regulate transcription by binding to specific DNA sequences known as motifs. TFs are involved in various biological processes, including immune response[4], maintenance of pluripotency of stem cells[5], and the dysfunctions of TFs are related to numerous human diseases, such as various types of cancer and neurodegenerative diseases[6,7]. Additionally, other DNA-binding proteins such as histone, DNA polymerase, and DNA topoisomerase, also play critical roles in biological activities and are associated with human diseases[8,9].

Identifying the DNA-binding sites of a protein is the initial step for modeling protein-DNA binding properties. Several experimental approaches have been developed for identifying protein-DNA interaction *in vivo* or *in vitro*, such as systematic evolution of ligands by exponential enrichment (SELEX) and chromatin immunoprecipitation (ChIP)[10,11]. In addition, structural biology approaches have been applied to determine the DNA-binding residues and areas, including X-ray crystallography and nuclear magnetic resonance (NMR). Although experimental methods based on molecular biology have made significant contributions over the past few decades, these methods are time-consuming and resource-intensive. Therefore, computationally predicting DNA-binding residues with machine learning methods is attractive.

The vital step in building a predictor is representation learning, where discriminative features play a crucial role in improving model performance. Typically, models utilize features extracted from a collection of protein sequences to fully leverage evolutionary information. The commonly used methods involve PSI-BLAST[12] and HHblits[13], which produce multiple sequence alignment (MSA) described as a position-specific scoring matrix (PSSM). Extensive studies show that evolutionary information leads to significant improvement in DNA-binding prediction tasks[14,15]. The secondary structure information of the given protein can also be applied as the initial feature, which can be generated by DSSP[16] using protein structure or PSIPRED[17] using protein sequence. A number of models have been developed to complete the task and can be roughly divided into sequence-based and structure-based models. Sequence-based models extract features from protein sequences alone, while structure-based models use features of crystal protein structures. BindN[18] used several amino acid properties as sequence features, applying a support vector machine (SVM) model to classify the DNA-binding residues, BindN+[14] improved the model performance by adding the PSSM feature. Currently, advanced predictors are focused on deep learning methods, with DeepDISE[19] and DBPred[15] using a convolutional neural network (CNN) as the classifier, EL_LSTM[20] applying a recurrent neural network (RNN) as the backbone network, and ProNA2020[21] using a multi-layer perceptron (MLP). A few models start with predicted protein structures or experimentally solved structures. NucBind[22] predicted protein structures by template-based models and then used an SVM to complete the downstream prediction. GraphBind integrated sequence-based and structure-based features, employing graph neural network (GNN) as the classifier.

Protein structures contain all the necessary information derived from the protein sequence. Hence, in general, structure-based models demonstrate better performance than sequence-based models. However, to ensure model performance, structure-based models require accurate protein structures as input. Consequently, the prediction of DNA-binding sites based on protein sequences remains an important and pressing research problem. Currently, the performance of existing sequence-based models is still unsatisfactory for practical application, and the feature extraction process often relies on manual design, which fails to generate a refined initial representation[23]. As a result, there is a pressing need to develop an end-to-end model that without using handcrafted features. Pre-training and contrastive learning are two widely-used representation learning techniques. Pre-training utilizes the information of a large scale of unlabeled data to train the model in an unsupervised manner and transfers the model parameters to downstream tasks for fine-tuning or feature extraction[24]. Contrastive learning aims to discover a representation space where samples from the same class are close to each other, while those from the different classes are distant.

In this study, we integrated pre-training and contrastive learning techniques to devise the CLAPE (Contrastive Learning And Pre-trained Encoder), which enabled the prediction of ligand-binding sites of a protein sequence. Specifically, we trained CLAPE-DB on DNA-binding datasets and demonstrated that it surpassed current sequence-based models by learning a discriminative embedding space. Additionally, we illustrated that CLAPE could serve as a general framework for predicting ligand-binding sites exclusively based on protein sequence information, thereby improving the comprehension of the feature extraction process and the development of the model architecture for future research.

## Results

**The model architecture of CLAPE**

The existing models for identifying protein-DNA binding sites could be divided into two categories. The first category combines handcrafted features and classification models (Figure 1a). Handcrafted features may include amino acid physicochemical properties and protein structural information, while the models may include machine learning models such as support vector machine and random forest. The second category aims to predict DNA-binding sites in an end-to-end fashion (Figure 1a) and often employs large-scale deep learning models. One may either train a classification model from scratch or fine-tune a pre-trained protein language model, such as ProtBert, with a simple downstream neural network such as linear layers. However, the first approach typically necessitates laborious manual feature extraction processes, while the second approach demands high computational resources and training time.

We took advantage of both approaches to propose CLAPE, a protein-ligand binding sites prediction framework to generate the binding probabilities of a given protein sequence. The overall architecture of CLAPE was depicted in Figure 1b, which comprised three main modules: the sequence embedding module, the backbone network module, and the loss computation module. The sequence embedding module utilized ProtBert[25], a pre-trained protein language model, to encode protein sequences in FASTA format and generate features with a dimensionality of 1024. The features were then passed through the backbone network, which in CLAPE was a 4-layer 1DCNN. The backbone network module generated a 2-dimensional matrix. The loss computation module employed a contrastive loss function, guided by binary classification loss, to optimize the model parameters. Finally, the classification head utilized a Softmax function to transform the prediction scores of the backbone network into the classification probabilities, which was common-used in current approaches.

CLAPE is a highly flexible framework, allowing for the customization of each basic component. In the loss computation module, one may employ different contrastive loss functions such as that proposed in DrLIM[26], triplet loss[27], or lifted structure loss[28]. While our experiments indicated that 1DCNN was the most effective backbone network for CLAPE, other backbone models, such as MLP (multi-layer perceptron) and RNN (recurrent neural network), were also suitable for use with CLAPE.

The pre-trained model was used as a feature extractor to avoid tedious manual feature extraction procedures. However, researchers may choose to fine-tune the pre-trained model, which has been shown to produce better performance but requires higher computational and time consumption[29] resembling the training scheme described in Figure 1a. Furthermore, multiple pre-trained protein language models can be applied to the sequence embedding module[30].

## CLAPE-DB accurately predicted the DNA-binding sites with a better generalization ability

We evaluated the performance of the proposed CLAPE-DB (CLAPE DNA-binding) model on two protein-DNA datasets, as described in Table 1. To assess the performance of CLAPE-DB, we conducted experiments on both Dataset1 and Dataset2 using independent testing set TE46 and TE129, respectively. We compared the results with existing DNA-binding sites prediction tools based on protein sequence input. CLAPE-DB outperformed other methods on both datasets (Table 2 and Table 3). Specifically, in TE46, CLAPE-DB trained on TR646 outperformed the second-best model DBPred[15] by a large margin, achieving a specificity of 0.835, a recall of 0.747, a precision of 0.306, an F1-score of 0.434, an MCC of 0.401, and an AUC of 0.871 in Dataset1 (Table 2), yielding a significant improvement over DBPred by 6.5%, 5.5%, 25.9%, 19.9%, 25.3%, 9.6%. Notably, DBPred used a manual feature extraction process, and a similar CNN model as CLAPE-DB, highlighting the advantages of using pre-trained models over handcrafted features.

Moreover, we trained and evaluated CLAPE-DB on Dataset2, and compared the performance with other existing tools (Table 3). CLAPE-DB also achieved a better predictive capability on this dataset, particularly in the recall metric, indicating the ability to accurately identify true DNA-binding sites. Furthermore, our proposed model demonstrated a significant improvement in the MCC metric over the previous models, suggesting a superior ability to handle imbalanced data. Dataset2 was proposed as a benchmark dataset for structure-based models, and we compared the metrics of several structure-based models (Supplementary Table 1). Although CLAPE-DB did not incorporate any structure information, it outperformed the structure-based models, such as COACH-D, NucBind, and DNAbind. Notably, the GraphBind model used predicted protein structure exhibited a poor performance with an AUC of 0.816, lower than that of CLAPE-DB. The results suggested that structure-based models required accurate protein structure to achieve acceptable prediction results. Moreover, compared to the structure-based models, CLAPE-DB used only a pre-trained language model and a simple backbone network to process the data, which reduced the model complexity and enhanced the inference speed, while maintaining accuracy. Our results also implied that pre-trained language models could capture some structural information from sequence inputs alone.

To test the generalization ability of CLAPE-DB, we trained a model on Dataset1 and tested it on Dataset2 (Figure 2a-b), as the protein sequence embeddings of Dataset1 and Dataset2 should have similar data distribution, DBPred was also tested for comparison. The prediction metrics of CLAPE-DB surpassed DBPred by a large margin, AUC and AUPR of CLAPE-DB were 0.865 and 0.394, respectively, while the metrics of DBPred standalone version were 0.526 and 0.068, which was slightly higher than a random choice result. Besides, the result of CLAPE-DB was merely lower than CLAPE-DB trained on TR573 (0.871 and 0.881 respectively), showing our proposed model had a superior generalization ability compared to the second-best sequence-based model, DBPred. For further clarification of the generalization ability of CLAPE-DB, we selected the dataset TE181 (Supplementary Table 2) created by Yuan et al.[31] and tested the performance of CLAPE-DB trained on TR573. CLAPE-DB still showed a better performance than other sequence-based models and most structure-based models (Supplementary Table 3, Supplementary Table 4). To be noted, the COACH-D and NucBind did not perform well on the TE181 dataset, showing the model performance was restricted to the limited accuracy of protein structure prediction based on homology modeling and molecular docking method.

**Backbone network comparison and feature visualization of CLAPE-DB**

We conducted experiments to compare the performance of different mainstream backbone networks, including MLP, RNN, and CNN, in predicting DNA-binding sites, and we used an LSTM model to represent the RNN model. 1DCNN model achieved the best performance among the three commonly-used models (Figure 2c). Although RNN was specifically designed for sequence modeling tasks, our finding suggested that the CNN was more suitable for predicting DNA-binding sites. This might be because RNN models process sequential data from left to right, whereas DNA-binding residues are predominately determined by spatial structures rather than simple sequential order. While CNN models the protein sequences using sliding windows, which incorporate relative positional information of amino acids inherently, amino acids are treated as independent tokens in RNN models[32].

To visualize the embedding space, we compared the embeddings generated by the CLAPE-DB model and an untrained, randomly initialized 1DCNN model, and we utilized t-SNE (t-distributed Stochastic Neighbor Embedding) dimension reduction method. Our results showed that CLAPE-DB learned a discriminative embedding space, while the data points were randomly distributed in the space after being processed by the untrained model (Figure 2d-e). Moreover, CLAPE-DB was able to effectively distinguish the DNA-binding and non-binding samples in the embedding space of each layer, with the distinction becoming more pronounced as the convolutional layer approached the output layer (Supplementary Figure 1). Additionally, we plotted the dimension reduction result of the raw features generated by ProtBert, which showed that the raw features were not well separated before model processing. Our results showed that CLAPE-DB was effective at distinguishing data samples from different classes (Figure 2f).

**Contrastive learning improved the model performance**

In the loss computation module, CLAPE-DB utilized a combination of triplet center loss (TCL)[34] and class-balanced focal loss[35,36]. To analyze the effectiveness of the loss functions, we performed ablation studies. TCL and focal loss generated discriminative embeddings in high-dimensional space, and both loss functions led to better performance than the commonly-used cross-entropy loss (Table 4). Furthermore, the improvement in the AUC value of the TCL function over the focal loss function was slightly smaller than that of cross-entropy loss (0.006 and 0.012, respectively), which might be attributed to the focal loss achieving better performance in modeling imbalanced datasets. The improvement in the AUPR value indicated that class-balanced focal loss and contrastive learning methods showed a better ability to cope with imbalanced datasets.

We also visualized the embeddings of the shape of 1024 generated by the first layer using focal loss only and jointly using focal loss and TCL. As expected, though the embeddings of DNA-binding sites and non-binding sites separated to a certain extent, the embeddings generated by joint loss functions showed a single clustering center, and the positive and negative samples were more discriminative (Supplementary Figure 2). The single and uniform cluster center could benefit the classification performance according to the previous studies[34,37].

**Parameter impact of loss functions**

The hyperparameters utilized in TCL and class-balanced loss matter in model training and inference, therefore, we analyzed and adjusted the hyperparameters in the loss functions. In the class-balanced focal loss, we implemented the effective number[36] as a reweight for samples from different classes, thereby we adjusted the hyperparameter $\gamma$. As $\gamma$ was an exponential parameter, an increase in the value would lead to a simultaneous reduction in the values of both hard and easy samples. Furthermore, this reduction would be more obvious in the case of hard samples[35]. We tested values of $\gamma$ from 1 to 10 and observed the AUC and AUPR values remained relatively stable within a specific range, but with an increase of $\gamma$, both metrics displayed a significant decline (Figure 3a). To verify our findings, we conducted further tests with $\gamma$ values of 0.5 and 20. Finally, we adopted a $\gamma$ value of 5. It is worth noting that as the total loss value decreased, a lower learning rate should be specified to ensure convergence.

During the optimization process of TCL, the cluster centers were randomly initialized, and we tested the model performance by adjusting the parameter learning rate and margin (m). Previous studies suggested that the learning rate for optimizing the cluster center should be relatively large[37]. However, we found that the AUC value was the highest when the learning rate was set to a relatively small value of 0.01 (Figure 3b). Margin was another crucial hyperparameter in TCL. If the margin was too large, the model might fail to recognize subtle differences between samples from different classes, and the convergence time might be prolonged. On the other hand, the loss of a lot of samples would be 0 when the margin was too small. Intuitively, we visualized the distance distribution of TR646 to guide our choice of parameter m. The distances from negative to positive and positive to negative were distributed from 7 to 12 (Figure 3c). Thus, we adjusted the margin value based on the distribution plot. We found that the AUC was maximized when the margin was set to 9, which was consistent with our expectations (Figure 3d). It is worth noting that the appropriate margin value was largely influenced by the data. Several attempts were made to adjust the margin according to the data distribution. For instance, Zhao et al[38] used the true distance to model the margin and Cheng et al[39] used a self-adaptive margin by a Gaussian prior distribution.

**CLAPE-DB captured the properties distribution of amino acids**

Based on previous studies, it is widely acknowledged that protein-DNA binding preferences are reflected in the sequences and structures of proteins and DNA[40]. For instance, proteins can bind DNA modules via hydrogen bonds and hydrophobic interactions. Such biological phenomena are related to the amino acid composition and properties of proteins. Therefore, in this study, we evaluated the predictive ability of CLAPE-DB using the TE129 dataset.

To this end, we performed a statistical analysis of the amino acid composition of DNA-binding sites and non-binding sites. Lysine, arginine, and tyrosine were the predominant amino acid types in the DNA binding sites, while alanine and leucine were the primary amino acid types in the non-binding sites (Figure 4a). Furthermore, we compared the amino acid type distribution of predicted results and the ground truth (Figure 4b) and used the Kullback-Leibler (KL) divergence to measure the distance of discrete distributions. The shapes of distributions of prediction and ground truth were found to be quite similar, and the forward and reverse KL divergence were 0.024 and 0.028, respectively, which were close to 0, indicating that the two distributions were semblable. Our results demonstrated that CLAPE-DB could accurately capture the amino acid composition features of DNA-binding residues.

Additionally, we analyzed the physicochemical properties of amino acids by extracting features from protein sequence and structure, and subsequently tested several selected properties, such as hydrophobicity, charge, secondary structure, and solvent accessibility. The t-SNE dimension reduction visualization revealed that different types of amino acid physiochemical properties were segregated into various clusters (Supplementary Figure 3a-d). Our results illustrated that the large-scale pre-trained protein language model ProtBert was capable of effectively learning the properties of amino acids. Such models were identified as appropriate feature extractors to replace handcrafted descriptors, which is congruent with previous studies[23].

Moreover, CLAPE-DB was proved successful in predicting not only the distribution of amino acids but also their properties. CLAPE-DB showed a similar distribution of different types of properties like the majority of amino acids were polar and positively charged. Furthermore, the binding sites predicted by CLAPE-DB exhibited a similar composition of different properties to the real DNA-binding sites (Figure 4c-f). Taken together, CLAPE-DB accurately captured the amino acid information that was analogous to real binding sites.

**Comparative and empirical case study**

To intuitively visualize and compare the prediction performance of DNA-binding residues of CLAPE-DB, we selected two protein structures for illustration purposes: multiple antibody resistance regulator (MarR) families (PDB ID: 5H3R, chain A, denoted as 5H3R_A) and transcription repressor protein CouR (PDB ID: 6C2S, chain A, denoted as 6C2S_A). CLAPE-DB made an accurate prediction of DNA-binding sites, while DBPred only captured a limited number of true positive sites, highlighting the superior prediction ability of CLAPE-DB. In addition, the majority of false positive sites were located in close proximity to binding sites (Figure 5a-f). Our results suggested that CLAPE-DB effectively learned the amino acid properties that were spatially adjacent and the structural information without relying on protein structures.

DNA molecules are negatively charged and tend to bind the positively charged regions of proteins. The structure of the protein-DNA binding area could be divided into several domains with specific patterns[41]. Empirical observations and computational properties can be utilized to infer the DNA-binding sites from the protein structure. However, such methods have significant limitations. Firstly, some proteins, such as intrinsically disordered proteins (IDP), are unstructured when not bound by ligands like DNA[42]. Therefore, the DNA-binding sites could not be inferred from the structure of such proteins. Secondly, the inferred probable DNA-binding sites using the surface charge distribution and protein structure are often quite different from the real binding sites. To illustrate the limitations of empirical analysis, we selected two protein structures: the transcription regulatory protein FadR (PDB ID: 5GPC, chain A, denoted as 5GPC_A) and bacteria quorum-sensing repressor protein RsaL (PDB ID: 5J2Y, chain A, denoted as 5J2Y_A). In both protein structures, multiple possible binding sites were identified based on the charge distribution (Figure 5g and Figure 5j), and it was difficult to determine which part of the protein would bind the major or minor groove of DNA. However, CLAPE-DB precisely distinguished the binding sites, and the false positive sites were not influenced by the other positively charged locations (Figure 5h-i and Figure 5k-l). It should be noted that the empirical binding site identification relied on the experimental structures, which was limited when lacking protein structures or using inaccurately predicted structures. Furthermore, some DNA-binding proteins, such as transcription activator-like effector nuclease (TALEN), are not typical in common empirical analyses. Therefore, precise DNA-binding site prediction using CLAPE-DB is necessary instead of relying on empirical inference.

**CLAPE was a general ligand-binding sites prediction framework**

CLAPE could serve as a general framework for predicting other ligand-binding sites, including protein-RNA and antibody-antigen binding sites. (Figure 6a-b). To evaluate the prediction ability of CLAPE for these types of binding sites, we collected benchmark datasets of protein-RNA and antibody-antigen binding sites (Supplementary Table 5), and trained CLAPE on these datasets. The resulting models were denoted as CLAPE-RB (CLAPE RNA-binding) and CLAPE-AB (CLAPE-Antibody). Both CLAPE-RB and CLAPE-AB performed well on the testing sets, with CLAPE-AB achieving the AUC of 0.920 (Supplementary Table 6), which was relatively high and could be applied to accurately predict the paratope of a given antibody sequence. It should be noted that the prediction capability of antigen-agnostic paratope was limited and could be improved by adding the epitope information[43]. The prediction task of RNA-binding sites was complicated due to the flexibility of the RNA structure, and the metrics of CLAPE-RB were relatively low compared to CLAPE-AB. Nevertheless, the AUC of CLAPE-RB trained on TE161 was 0.830 (Supplementary Table 6), which surpassed the existing sequence-based RNA-binding sites models[44,45]. We also plotted the ROC and AUC curves to visualize the overall model performance of CLAPE-RB and CLAPE-AB (Figure 6c-d).

To evaluate the performance of our model, we trained CLAPE-RB on a separate protein-RNA dataset comprising TR495 and TE117, which were widely used benchmarks for structure-based models. CLAPE-RB outperformed existing sequence-based models in predicting RNA-binding sites on TE117. While the performance of CLAPE-RB was marginally lower than that of the structure-based model GraphBind, it performed better than Nucleic, a CNN model predicting RNA binding sites based on grids of the protein surface (Supplementary Table 8). Similarly, CLAPE-RB outperformed GraphBind based on inaccurately predicted protein structure, which highlighted the potential of CLAPE to overcome the limitations of structure-based models. Our results indicated that CLAPE was a versatile framework that could predict ligand-binding sites of a given protein sequence for a range of ligands. Furthermore, our experiments demonstrated that CLAPE, based on a large-scale pre-trained language model, was an effective predictor of ligand-binding sites, even in the absence of structural information, achieving relatively high performance.

**CLAPE-DB exclusively predicted DNA-binding sites**

Previous studies demonstrated that different ligands tend to bind to different sites on proteins[46], implying that the performance of the CLAPE model trained on specific ligands, such as DNA, may be inferior compared to other types of ligands. To validate this hypothesis, we evaluated the performance of CLAPE-DB on RNA-binding sites prediction (using TE117) and antibody-antigen binding sites prediction (using TE259) tasks.

The various metrics of CLAPE-DB on DNA-binding sites, including precision, recall, F1-score, and MCC, were much higher than those of other ligand-binding sites (Supplementary Figure 4a), indicating CLAPE-DB was a specific predictor for DNA-binding sites. Notably, CLAPE-DB showed a poor ability to predict paratopes, which was consistent with the unique characteristics of antibody protein sequences. Furthermore, according to the predictive results (Supplementary Table 6), CLAPE-AB achieved an AUC value greater than 0.9, providing further evidence that CLAPE-DB learned discriminative features of DNA-binding protein sequences and that CLAPE was a general predictor for various ligand-binding sites.

We also plotted the ROC curve of the CLAPE-DB model on different ligands (Supplementary Figure 4b), and the AUC value of CLAPE-DB on DNA was still higher than on other binding sites. Notably, the AUC of CLAPE-DB on RNA reached 0.775, which was comparable to the performance of existing models trained on RNA datasets, such as RNABindPlus, SVMnuc, and GraphBind based on predicted protein structures. Our results suggested that RNA and DNA had similarities in terms of their protein binding patterns, given their nucleic-acid nature. However, RNA molecules are typically single-chain and have more complicated conformations, which increases the difficulty of prediction.

## Discussion

Protein-DNA binding plays an important role in many life activities, and studies on the binding properties contribute to the understanding of genome transcription and regulation. Accurate identification of DNA-binding sites of proteins is a crucial step in modeling the protein-DNA interactions. Various models have been developed using machine learning and deep learning techniques to identify DNA-binding sites from protein sequence or structure[15,47]. However, current tools rely on tedious manual feature extraction processing, which is time-consuming and redundant. Additionally, the accuracy of sequence-based models still needs to be increased, and the performance of the structure-based models is largely affected by the accuracy of protein structure, restricting their widespread application. Given these limitations, it is imperative to develop a satisfactory sequence-based model that utilizes protein sequence information alone to predict DNA-binding sites. To address the existing challenges and improve the performance of the sequence-based models, we proposed CLAPE, a deep learning framework that combines a large-scale pre-trained protein language model and contrastive learning technique to accurately predict DNA-binding sites of a given protein sequence. We performed multiple experiments to evaluate the performance and effectiveness of our proposed model.

In this study, we presented the overall architecture of CLAPE, which was comprised of three main components. Firstly, we utilized a pre-trained model, ProtBert, without fine-tuning, to conduct feature extraction. Secondly, we employed a 1DCNN to process the sequence feature and generate the classification score. Finally, we jointly optimized a class-balanced focal loss and a contrastive triplet center loss to address the issue of imbalanced data, which resulted in a more discriminative embedding space with a single cluster center.

The proposed CLAPE-DB model for predicting DNA-binding sites demonstrated superior performance compared to existing sequence-based models on two benchmark datasets, as indicated by all metrics, with an AUC of 0.871 and 0.881, respectively. Furthermore, in cases where accurate protein crystal structures were unavailable, CLAPE-DB outperformed structure-based models by a large margin. Additionally, we evaluated the generalization ability of the CLAPE-DB model on independent datasets and found that CLAPE-DB exhibited better generalization performance than the second-best model, DBPred. These results suggested that CLAPE-DB effectively learned the underlying latent distribution of DNA-binding sites.

To mitigate the effects of imbalanced data, we implemented the class-balanced focal loss in our proposed CLAPE model. There were several augmentation approaches from the aspect of the dataset, such as SMOTE[48] (synthetic minority over-sampling technique) to interpolate new data in the embedding space. The sequence and structure alignment methods could be used to transfer the annotation of the known DNA-binding sites of the matched proteins. Furthermore, incorporating the newly solved protein-DNA complexes into the dataset could enhance the prediction performance and generalization ability of the model. Additionally, various contrastive loss functions, such as the lifted structure loss and N-pair loss, could be employed for these tasks. The lifted structure loss considered all negative samples of the batch in a single optimization procedure, while N-pair loss included pairs from all classes as negative samples.

CLAPE-DB showed a more discriminative embedding space via the visualization of dimension reduction of the hidden layers. In addition, the feature generated by ProtBert could capture the amino acid physicochemical properties and distributions. Our study demonstrated that a large-scale pre-trained protein language model could extract protein sequence features effectively, eliminating the need for designing handcrafted features. In this study, we only evaluated the ProtBert as the feature extractor, but other pre-trained protein models such as RITA[49] and ESM-2[50], as reviewed in detail by Hu et al[30]. could also be used for feature generation. Here, we tested the performance of CLAPE-DB applying a larger protein language model ESM-2 as the feature extractor, which contained more parameters than ProtBert, and the model performance of CLAPE-DB was improved using ESM-2 which was consistent with our expectation (Supplementary Table 9).

In the computer vision and natural language processing fields, there is a trend toward utilizing a unified large model for addressing multiple downstream tasks[51], namely artificial general intelligence (AGI), which could also be applied to generate embeddings of protein sequences. Although the CLAPE-DB was designed for sequence-based prediction tasks, it's possible to use the features generated by the pre-trained model for the structure-based model, as demonstrated in related studies[52].

Based on the results of our experiments, we conclude that CLAPE is a general prediction framework for identifying ligand-binding sites of a given protein sequence. In addition, CLAPE-RB and CLAPE-AB demonstrated satisfactory performance on their respective datasets. Moreover, we showed that CLAPE-DB could exclusively predict the DNA-binding sites, which was not generally influenced by the information from other ligand-binding sites. The simple 1DCNN model used in all CLAPE series models effectively captured the neighboring information of the targeted residue. Given the flexibility of CLAPE, various backbone models could be applied, such as attention-based models for modeling the long-range relationship of residues[53].

Overall, the deep learning model CLAPE proposed in our study achieved high performances in predicting both DNA- and ligand-binding sites by combining pre-trained models with contrastive learning methods. The promising and general framework can be applied in future studies to facilitate protein function annotation, protein engineering, and drug discovery.

## Methods

**Datasets description**

In this study, we evaluated and compared the performance of our proposed model, CLAPE, with existing classifiers using two widely used benchmark datasets, denoted as Dataset1 and Dataset2. The training and testing datasets were denoted as TR and TE, respectively. Both datasets were preprocessed by similar procedures to improve the robustness of models and avoid bias due to the imbalanced data distribution, such as reducing the sequence similarity using a cutoff of 30% with CD-HIT[54]. The binding sites were defined similarly in both datasets as residues with a distance less than 0.5 plus the sum of the Van der Waals radius of the two nearest atoms between the residue and the nucleic acid molecule. Table 1 provides a summary of the benchmark datasets, and the details of both datasets are described below.

Dataset1 was introduced by the study of the DBPred model, a sequence-based deep learning method for predicting DNA-binding residues[15]. The dataset was collected from hybridNAP[55] and ProNA2020[21] and was composed of 646 proteins as the training set (TR646) with 15636 DNA-binding sites and 298503 non-binding sites, and 46 proteins as the testing set (TE46) with 956 DNA-binding sites and 9911 non-binding sites.

Dataset2 was originally proposed by the study of GraphBind, a structure-based graph neural network (GNN) model for identifying nucleic-acid-binding residues[47]. This dataset consisted of protein-DNA complex structural data extracted from the BioLiP database[56], with 573 proteins as a training set (TR573) with 14479 DNA-binding residues and 145404 non-binding residues, and 129 proteins as a testing set (TE129) with 2240 DNA-binding residues and 35275 non-binding residues. GraphBind employed a data augmentation approach on the training set to alleviate the impact of the data imbalanced issue, hence we used the same augmented data annotations as GraphBind.

To assess the prediction capability of our proposed model CLAPE on diverse ligand-binding sites, we gathered three different datasets comprising protein-RNA and antibody-antigen interactions. The protein-RNA datasets were created by Xia et al. based on the GraphBind model, and Patiyal et al., based on the pprint2 model[45]. The antibody-antigen dataset was collected from the SAbDab database[57]. To ensure a fair comparison with existing models, we applied the same data preprocessing procedure as used for defining DNA-binding sites.

**Protein sequence embedding**

The protein sequences were first input into ProtBert[25], a pre-trained model, to generate high-dimensional embeddings. ProtBert is a member of the ProtTrans family of pre-trained models and is based on the BERT architecture. The ProtTrans models were trained on large-scale protein sequences and have been commonly used for predicting protein structure and properties. BERT employed a masked language modeling strategy to train a Transformer encoder[58,59], which could effectively embed target tokens with contextual information. This approach is often used in token-level NLP tasks such as named entity recognition (NER). Since the task of predicting DNA-binding residues in protein sequences was also a token-level classification task, we utilized ProtBert, which has 400 million parameters, as a feature extractor. The dimension of the protein embedding generated by ProtBert was 1024. It is important to note that ProtBert was not fine-tuned during subsequent training steps, and the sequence embedding process was performed using HuggingFace's Transformers Python package[60].

**Backbone 1DCNN model and classification head**

We utilized a 1DCNN (one-dimensional convolutional neural network) as our backbone model to obtain a residue-level classification score. The convolutional neural network used a convolutional kernel to capture neighboring information and used operations like max pooling for down-sampling. In a single step of 1DCNN operation, the shape of input and output features were $[B, L, D_{i-1}]$ and $[B, L, D_i]$, respectively, where $B$ stood for the batch size of input data, $L$ was the maximum length of the protein sequence, $D_{i-1}$ and $D_i$ were the dimension of the last layer and the current layer, respectively. To maintain the same length of input and output protein sequence and obtain a unified token-level classification result, we applied padding for different convolutional kernel sizes. The stride of every layer was set to 1, and we utilized ReLU (rectified linear unit) as an activation function to introduce nonlinearity to the model. We applied dropout and batch normalization techniques to enhance the robustness and generalization ability of the model. Our CLAPE-DB model consisted of 4 1DCNN layers as the backbone model. The raw dimension was 1024, and the output dimension of the 4 layers were 1024, 128, 64, and 2, respectively. The classification head part contained a Softmax function to scale the output value between 0-1 as a mutually exclusive prediction score, representing the classification probability of DNA-binding sites.

**Binary classification loss function**

The classification loss function is crucial in neural network design as it measures the difference between predicted and true labels. Cross entropy is a commonly used loss function in binary classification tasks. However, protein-DNA binding data confronted a data imbalance issue as shown in Table 1, thus we applied a class-balanced focal loss to address this problem.

The focal loss was introduced by Lin et al.[35] and places more emphasis on classes with fewer samples in the loss function. It also considers the difficulty of samples based on the classification probability provided by the classifier. Specifically, if the classification probability was high enough, the sample would be defined as an easy sample. The focal loss is formulated as follows:

$$FL(p_t) = -\alpha_t(1-p_t)^\gamma log(p_t) \quad (1),$$

where $p_t$ is the classification probability of a particular class, $1-p_t$ is the modulator, $\gamma$ is a hyperparameter to adjust the weight of hard and easy samples. In the original paper, $\alpha$ is also a parameter to give the weight of minority and majority samples, which is influenced by $\gamma$. We applied an effective number to reweight the focal loss which was proposed by Cui et al[36], the basic hypothesis behind this idea was, with the increasing number of samples, the overlapping of embeddings would lead to information redundancy, thus effective number was proposed to model the real space covered by all samples, which could be used as a weight for imbalanced data. The class-balanced focal loss can be formulated as:

$$L_{focal} = -\frac{1-\beta}{1-\beta^{n_y}}\sum_{i=1}^{C}(1-p_i^t)^\gamma log(p_i^t) \quad (2),$$

[61]where $E_n = (1-\beta^n)/(1-\beta)$ refers to the effective number of the class, we set $\beta$ to 0.999 in our study according to Cui et al[36]. The class-balanced focal loss was jointly optimized with contrastive loss, as described in the following parts.

**Contrastive learning loss**

Contrastive learning aims to identify an embedding space where similar samples are positioned close to each other, while the dissimilar ones are far apart. Contrastive learning techniques have been extensively used in computer vision and natural language processing, and several models have shown promising results in representation learning, such as MoCo[62] and SimSiam[61]. In our study, we applied a contrastive loss, namely triplet center loss (TCL)[34], which is a supervised approach that takes into account the labels of the training data. TCL is a combination of center loss[37] and triplet loss[27], which

could be described as:

$$L_{tpl} = \sum_{i=1}^{N} max\left(0, m + D\left(f(x_a^i), f(x_+^i)\right) - D\left(f(x_a^i), f(x_-^i)\right)\right) \quad (3),$$

$$L_c = \frac{1}{2}\sum_{i=1}^{n} D(f(x^i), c_{y_i}) \quad (4),$$

where $m$ is the margin value, the expected distance of a given positive and negative sample, $c_{y_i}$ is the center of the given class $y_i$, and $x_a$ is an anchor point, while $x_+$ and $x_-$ are positive and negative data samples, respectively. TCL makes the positive and negative samples far away from each other and forces the samples of different classes to be close to respective cluster centers. The formulation of TCL can be mathematically expressed as follows:

$$L_{tc} = \sum_{i=1}^{M} max\left(D(f_i, c_{y^i}) + m - \min_{j \neq y^i} D(f_i, c_j), 0\right) \quad (5),$$

the positive and negative sample pair is substituted by the clustering center of samples, and $f_i$ refers to the classification probability predicted by the model. $D$ indicates the Euclidean distance between data points: $D(f_i, c_{y^i}) = \frac{1}{2}|f_i - c_{y^i}|_2^2$. The total loss was weighted by class-balanced focal loss and TCL using a hyperparameter $\lambda$, which was set to 0.1 in our study after searching, the loss function could be formulated as follows:

$$L = L_{focal} + \lambda L_{tc} \quad (6).$$

The backpropagation stops at the embedding generated by ProtBert, which means we did not fine-tune the pre-trained language model.

**Evaluation metrics**

In this study, we employed several classification evaluation metrics to ensure consistency with the previous studies. The metrics included specificity (Spe), precision (Pre), recall (Rec), F1-score, and Matthews correlation coefficient (MCC). The metrics can be formulated as follows:

$$Spe = \frac{TN}{TN+FP} \quad (7),$$

$$Pre = \frac{TP}{TP+FP} \tag{8}$$

$$Rec = \frac{TP}{TP+FN} \tag{9}$$

$$F1 = 2 \times \frac{Pre \times Rec}{Pre+Rec} \tag{10}$$

$$MCC = \frac{TP \times TN - FN \times FP}{\sqrt{(TP+FP) \times (TP+FN) \times (TN+FP) \times (TN+FN)}} \tag{11}$$

where TP, FP, TN, and FN stand for true positive (number of residues that are correctly classified as DNA-binding sites), false positive (number of residues that are incorrectly classified as DNA-binding sites), true negative (number of residues that correctly classified as non-binding sites) and false negative (number of residues that incorrectly classified as non-binding sites), respectively. Specifically, specificity indicates the portion of correctly predicted non-binding sites, precision measures the accuracy of residues predicted as DNA-binding sites, recall measures the portion of DNA-binding residues successfully discovered by the model, and F1-score is the harmonic mean of precision and recall. MCC evaluates the prediction ability of both positive and negative classes of the model and is commonly used in imbalanced data. Besides, we plotted the ROC (receiver operating characteristic) curve and precision-recall curve to illustrate the overall performance of a model and used two threshold-agnostic metrics AUC (area under ROC curve) and AUPR (area under PR curve) as numerical evaluations of both curves.

## Data and code availability

The datasets of our study and the codes of CLAPE are freely available at https://github.com/YAndrewL/clape.

## Acknowledgments

This work was supported by the Tsinghua University Initiative Scientific Research Program (No. 20221080025) and the Tsinghua-Peking University Center for Life

# References

1  Dillon, S. C. & Dorman, C. J. Bacterial nucleoid-associated proteins, nucleoid structure and gene expression. *Nature Reviews Microbiology* 8, 185-195 (2010).

2  Lambert, S. A. *et al.* The human transcription factors. *Cell* 172, 650-665 (2018).

3  Walter, M. C. *et al.* PEDANT covers all complete RefSeq genomes. *Nucleic acids research* 37, D408-D411 (2009).

4  Ono, M. *et al.* Foxp3 controls regulatory T-cell function by interacting with AML1/Runx1. *Nature* 446, 685-689 (2007).

5  Takahashi, K. *et al.* Induction of pluripotent stem cells from adult human fibroblasts by defined factors. *cell* 131, 861-872 (2007).

6  Lu, T. *et al.* REST and stress resistance in ageing and Alzheimer's disease. *Nature* 507, 448-454 (2014).

7  Kawamura, M. *et al.* Loss of nuclear REST/NRSF in aged-dopaminergic neurons in Parkinson's disease patients. *Neuroscience Letters* 699, 59-63 (2019).

8  Liu, Z. *et al.* Drug discovery targeting bromodomain-containing protein 4. *Journal of medicinal chemistry* 60, 4533-4558 (2017).

9  Ratti, A. & Buratti, E. Physiological functions and pathobiology of TDP-43 and FUS/TLS proteins. *Journal of neurochemistry* 138, 95-111 (2016).

10  Furey, T. S. ChIP–seq and beyond: new and improved methodologies to detect and characterize protein–DNA interactions. *Nature Reviews Genetics* 13, 840-852 (2012).

11  Ferraz, R. A. C. *et al.* DNA–protein interaction studies: a historical and comparative analysis. *Plant Methods* 17, 1-21 (2021).

12  Neuwald, A. F. & Poleksic, A. PSI-BLAST searches using hidden markov models of structural repeats: prediction of an unusual sliding DNA clamp and of β-propellers in UV-damaged DNA-binding protein. *Nucleic acids research* 28, 3570-3580 (2000).

13  Remmert, M., Biegert, A., Hauser, A. & Söding, J. HHblits: lightning-fast iterative protein sequence searching by HMM-HMM alignment. *Nature methods* 9, 173-175 (2012).

14  Wang, L., Huang, C., Yang, M. Q. & Yang, J. Y. BindN+ for accurate prediction of DNA and RNA-binding residues from protein sequence features. *BMC Systems Biology* 4, 1-9 (2010).

15  Patiyal, S., Dhall, A. & Raghava, G. P. A deep learning-based method for the prediction of DNA interacting residues in a protein. *Briefings in Bioinformatics* 23 (2022).

16  Joosten, R. P. *et al.* A series of PDB related databases for everyday needs. *Nucleic acids research* 39, D411-D419 (2010).

17  McGuffin, L. J., Bryson, K. & Jones, D. T. The PSIPRED protein structure prediction server. *Bioinformatics* 16, 404-405 (2000).

18  Wang, L. & Brown, S. J. BindN: a web-based tool for efficient prediction of DNA and RNA binding sites in amino acid sequences. *Nucleic acids research* 34, W243-W248 (2006).

19  Hendrix, S. G., Chang, K. Y., Ryu, Z. & Xie, Z.-R. DeepDISE: DNA Binding Site Prediction Using a Deep Learning Method. *International Journal of Molecular Sciences* 22, 5510 (2021).

20  Zhou, J., Lu, Q., Xu, R., Gui, L. & Wang, H. EL_LSTM: prediction of DNA-binding residue from protein sequence by combining long short-term memory and ensemble learning. *IEEE/ACM transactions on computational biology and bioinformatics* 17, 124-135 (2018).

21  Qiu, J. *et al.* ProNA2020 predicts protein–DNA, protein–RNA, and protein–protein binding proteins and residues from sequence. *Journal of molecular biology* 432, 2428-2443 (2020).

22  Su, H., Liu, M., Sun, S., Peng, Z. & Yang, J. Improving the prediction of protein–nucleic acids binding residues via multiple sequence profiles and the consensus of complementary methods. *Bioinformatics* 35, 930-936 (2019).

23  Villegas-Morcillo, A. *et al.* Unsupervised protein embeddings outperform hand-crafted sequence and structure features at predicting molecular function. *Bioinformatics* 37, 162-170 (2021).

24  Han, X. *et al.* Pre-trained models: Past, present and future. *AI Open* 2, 225-250 (2021).

25  Elnaggar, A. *et al.* Prottrans: Toward understanding the language of life through self-supervised learning. *IEEE transactions on pattern analysis and machine intelligence* 44, 7112-7127 (2021).

26  Hadsell, R., Chopra, S. & LeCun, Y. Dimensionality Reduction by Learning an Invariant Mapping. *2006 IEEE Computer Society Conference on Computer Vision and Pattern Recognition (CVPR'06)* 2, 1735-1742 (2006).

27  Schroff, F., Kalenichenko, D. & Philbin, J. FaceNet: A unified embedding for face recognition and clustering. *2015 IEEE Conference on Computer Vision and Pattern Recognition (CVPR)*, 815-823 (2015).

28  Song, H. O., Xiang, Y., Jegelka, S. & Savarese, S. Deep Metric Learning via Lifted Structured Feature Embedding. *2016 IEEE Conference on Computer Vision and Pattern Recognition (CVPR)*, 4004-4012 (2015).

29  Wang, R., Jin, J., Zou, Q., Nakai, K. & Wei, L. Predicting protein-peptide binding residues via interpretable deep learning. *Bioinformatics* (2022).

30  Hu, B. *et al.* Protein Language Models and Structure Prediction: Connection and Progression. *arXiv preprint arXiv:2211.16742* (2022).

31  Yuan, Q. *et al.* AlphaFold2-aware protein-DNA binding site prediction using graph transformer. *Briefings in bioinformatics* (2022).

32  Yang, K. K., Fusi, N. & Lu, A. X. Convolutions are competitive with transformers for protein sequence pretraining. *bioRxiv*, 2022.2005. 2019.492714 (2022).

33  Ying, C. *et al.* Do transformers really perform badly for graph representation? *Advances in Neural Information Processing Systems* 34, 28877-28888 (2021).

34  He, X., Zhou, Y., Zhou, Z., Bai, S. & Bai, X. Triplet-Center Loss for Multi-view 3D Object Retrieval. *2018 IEEE/CVF Conference on Computer Vision and Pattern Recognition*, 1945-1954 (2018).

35  Lin, T.-Y., Goyal, P., Girshick, R. B., He, K. & Dollár, P. Focal Loss for Dense Object Detection. *IEEE Transactions on Pattern Analysis and Machine Intelligence* 42, 318-327 (2017).

36  Cui, Y., Jia, M., Lin, T.-Y., Song, Y. & Belongie, S. J. Class-Balanced Loss Based on Effective Number of Samples. *2019 IEEE/CVF Conference on Computer Vision and Pattern Recognition (CVPR)*, 9260-9269 (2019).

37   Wen, Y., Zhang, K., Li, Z. & Qiao, Y. A Discriminative Feature Learning Approach for Deep Face Recognition. in *European Conference on Computer Vision.* 2016:499-515.

38   Zhao, X., Qi, H., Luo, R. & Davis, L. A Weakly Supervised Adaptive Triplet Loss for Deep Metric Learning. *2019 IEEE/CVF International Conference on Computer Vision Workshop (ICCVW)*, 3177-3180 (2019).

39   Cheng, R., Wang, F., Zhao, T., Liu, H. & Zeng, H. Soft Margin Triplet-Center Loss for Multi-View 3D Shape Retrieval. *Int. J. Pattern Recognit. Artif. Intell.* 36, 2250017:2250011-2250017:2250019 (2022).

40   Rohs, R. *et al.* Origins of specificity in protein-DNA recognition. *Annual review of biochemistry* 79, 233-269 (2010).

41   Luscombe, N. M., Austin, S. E., Berman, H. M. & Thornton, J. M. An overview of the structures of protein-DNA complexes. *Genome Biology* 1, reviews001.001 - reviews001.037 (2000).

42   Sandhu, K. S. & Dash, D. Dynamic α-helices: Conformations that do not conform. *Proteins: Structure* 68 (2007).

43   Akbar, R. *et al.* Progress and challenges for the machine learning-based design of fit-for-purpose monoclonal antibodies. *mAbs* 14 (2022).

44   Wang, N., Yan, K., Zhang, J. & Liu, B. iDRNA-ITF: identifying DNA-and RNA-binding residues in proteins based on induction and transfer framework. *Briefings in Bioinformatics* 23, bbac236 (2022).

45   Patiyal, S., Dhall, A., Bajaj, K., Sahu, H. & Raghava, G. P. Prediction of RNA-interacting residues in a protein using CNN and evolutionary profile. *Briefings in Bioinformatics* 24, bbac538 (2023).

46   Zhang, J. & Kurgan, L. Review and comparative assessment of sequence-based predictors of protein-binding residues. *Briefings in Bioinformatics* 19, 821–837 (2018).

47   Xia, Y., Xia, C.-Q., Pan, X. & Shen, H.-B. GraphBind: protein structural context embedded rules learned by hierarchical graph neural networks for recognizing nucleic-acid-binding residues. *Nucleic acids research* 49, e51-e51 (2021).

48   Chawla, N. V., Bowyer, K. W., Hall, L. O. & Kegelmeyer, W. P. SMOTE: synthetic minority over-sampling technique. *Journal of artificial intelligence research* 16, 321-357 (2002).

49   Hesslow, D., Zanichelli, N., Notin, P., Poli, I. & Marks, D. Rita: a study on scaling up generative protein sequence models. *arXiv preprint arXiv:2205.05789* (2022).

50   Lin, Z. *et al.* Evolutionary-scale prediction of atomic-level protein structure with a language model. *Science* 379, 1123-1130 (2023).

51   Brown, T. *et al.* Language models are few-shot learners. *Advances in neural information processing systems* 33, 1877-1901 (2020).

52   Abdin, O., Nim, S., Wen, H. & Kim, P. M. PepNN: a deep attention model for the identification of peptide binding sites. *Communications Biology* 5, 503 (2022).

53   Lin, T., Wang, Y., Liu, X. & Qiu, X. A survey of transformers. *AI Open* (2022).

54   Li, W. & Godzik, A. Cd-hit: a fast program for clustering and comparing large sets of protein or nucleotide sequences. *Bioinformatics* 22 13, 1658-1659 (2006).

55 Zhang, J., Ma, Z. & Kurgan, L. Comprehensive review and empirical analysis of hallmarks of DNA-, RNA- and protein-binding residues in protein chains. *Briefings in bioinformatics* (2019).

56 Yang, J., Roy, A. & Zhang, Y. BioLiP: a semi-manually curated database for biologically relevant ligand–protein interactions. *Nucleic Acids Research* 41, D1096 - D1103 (2012).

57 Dunbar, J. *et al.* SAbDab: the structural antibody database. *Nucleic Acids Research* 42, D1140 - D1146 (2013).

58 Vaswani, A. *et al.* Attention is all you need. *Advances in neural information processing systems* 30 (2017).

59 Devlin, J., Chang, M.-W., Lee, K. & Toutanova, K. Bert: Pre-training of deep bidirectional transformers for language understanding. *arXiv preprint arXiv:1810.04805* (2018).

60 Wolf, T. *et al.* HuggingFace's Transformers: State-of-the-art Natural Language Processing. *ArXiv* abs/1910.03771 (2019).

61 Chen, X. & He, K. Exploring simple siamese representation learning. in *Proceedings of the IEEE/CVF conference on computer vision and pattern recognition.* 2021: 15750-15758.

62 He, K., Fan, H., Wu, Y., Xie, S. & Girshick, R. Momentum Contrast for Unsupervised Visual Representation Learning. in *Proceedings of the IEEE/CVF conference on computer vision and pattern recognition.* 2020: 9729-9738.

# Tables

Table 1: Summary of benchmark protein-DNA binding datasets

| Datasets | Dataset1 | | Dataset2 | |
|---|---|---|---|---|
| | TR646 | TE46 | TR573 | TE129 |
| DNA-binding residues | 15636 | 965 | 14479 | 2240 |
| Non-binding residues | 298503 | 9911 | 145404 | 35275 |
| % of binding residues | 4.98 | 8.87 | 9.06 | 5.97 |

Table 2: Comparison of CLAPE-DB with other sequence-based methods on TE46

| Models | Spe | Rec | Pre | F1 | MCC | AUC |
|---|---|---|---|---|---|---|
| DRNAPred | 0.692 | 0.677 | 0.185 | 0.291 | 0.226 | 0.755 |
| DNAPred | 0.655 | 0.671 | 0.157 | 0.254 | 0.194 | 0.730 |
| SVMnuc | 0.666 | 0.668 | 0.154 | 0.250 | 0.192 | 0.715 |
| NCBRPred | 0.674 | 0.677 | 0.165 | 0.265 | 0.207 | 0.713 |
| DBPred | 0.784 | 0.708 | 0.243 | 0.362 | 0.320 | 0.794 |
| **CLAPE-DB** | **0.835** | **0.747** | **0.306** | **0.434** | **0.401** | **0.871** |

Table 3: Comparison of CLAPE-DB with other sequence-based methods on TE129

| Models | Spe | Rec | Pre | F1 | MCC | AUC |
|---|---|---|---|---|---|---|
| DRNAPred | 0.937 | 0.233 | 0.190 | 0.210 | 0.155 | 0.693 |
| DNAPred | 0.954 | 0.396 | 0.353 | 0.373 | 0.332 | 0.845 |
| SVMnuc | 0.966 | 0.316 | 0.371 | 0.341 | 0.304 | 0.812 |
| NCBRPred | **0.969** | 0.312 | 0.392 | 0.347 | 0.313 | 0.823 |
| **CLAPE-DB** | 0.955 | **0.464** | **0.396** | **0.427** | **0.389** | **0.881** |

Table 4: Model performance using different loss functions

| Loss functions | AUC | AUPR |
| --- | --- | --- |
| Cross Entropy | 0.849 | 0.438 |
| Cross Entropy + TCL | 0.861 | 0.445 |
| Focal Loss | 0.865 | 0.459 |
| **Focal Loss + TCL** | **0.871** | **0.463** |

**Supplementary Tables**

Supplementary Table 1: Comparison of structure-based model with other methods on TE129

| Models | Spe | Rec | Pre | F1 | MCC | AUC |
| --- | --- | --- | --- | --- | --- | --- |
| COACH-D | 0.958 | 0.367 | 0.357 | 0.362 | 0.321 | 0.710 |
| NucBind | 0.966 | 0.330 | 0.381 | 0.354 | 0.317 | 0.811 |
| DNABind | 0.926 | 0.601 | 0.346 | 0.440 | 0.411 | 0.858 |
| GraphBind[a] | - | 0.439 | 0.310 | 0.362 | 0.320 | 0.816 |
| **GraphBind[b]** | **0.941** | **0.684** | **0.422** | **0.522** | **0.500** | **0.928** |

[a] indicates results using predicted protein structure [b] indicates results using experimental protein structures. Results were obtained from Xia et al[47]

Supplementary Table 2: Summary of protein-DNA binding sites dataset TE181

| Dataset | TE181 |
| --- | --- |
| DNA-binding residues | 3208 |
| Non-binding residues | 72050 |
| % of binding residues | 4.26 |

Supplementary Table 3: Comparison of CLAPE-DB with other methods on TE181

| Models | Spe | Rec | Pre | F1 | MCC | AUC |
|---|---|---|---|---|---|---|
| DNAPred | 0.948 | 0.334 | 0.223 | 0.267 | 0.233 | 0.802 |
| SVMnuc | 0.960 | 0.289 | 0.242 | 0.263 | 0.229 | 0.803 |
| NCBRPred | **0.964** | 0.259 | 0.241 | 0.250 | 0.215 | 0.771 |
| **CLAPE-DB** | 0.931 | **0.413** | 0.212 | **0.280** | **0.252** | **0.824** |

Supplementary Table 4: Comparison of structure-based model with other methods on TE181

| Models | Spe | Rec | Pre | F1 | MCC | AUC |
|---|---|---|---|---|---|---|
| COACH-D | **0.971** | 0.254 | 0.280 | 0.266 | 0.235 | 0.655 |
| NucBind | 0.960 | 0.293 | 0.248 | 0.269 | 0.234 | 0.796 |
| DNABind | 0.904 | 0.535 | 0.199 | 0.290 | 0.279 | 0.825 |
| GraphBind | 0.933 | **0.624** | 0.293 | 0.399 | 0.392 | 0.904 |
| **GraphSites** | 0.958 | 0.517 | **0.354** | **0.420** | **0.397** | **0.917** |

Supplementary Table 5: Summary of protein-ligand binding sites datasets

| Datasets | Protein-RNA | | Antibody-antigen | |
|---|---|---|---|---|
| | TR545 | TE161 | TE1011 | TE259 |
| Ligand-binding residues | 18559 | 6966 | 15749 | 3755 |
| Non-binding residues | 171879 | 44349 | 189519 | 50890 |
| % of binding residues | 9.75 | 13.58 | 7.67 | 6.87 |

Supplementary Table 6: The performance of CLAPE for other ligand-binding sites prediction.

| Models | F1 | MCC | AUC | AUPR |
|---|---|---|---|---|
| CLAPE-RB | 0.495 | 0.407 | 0.830 | 0.511 |
| CLAPE-AB | 0.567 | 0.534 | 0.920 | 0.568 |

Supplementary Table 7: Summary of protein-RNA binding sites datasets TR495 and TE117.

| Dataset | TR495 | TE117 |
|---|---|---|
| DNA-binding residues | 14609 | 2031 |
| Non-binding residues | 122290 | 35314 |
| % of binding residues | 10.76 | 5.44 |

Supplementary Table 8: Comparison of CLAPE-RB with other RNA-biding sites prediction tools on TE117

| Models | Rec | Pre | F1 | MCC | AUC |
|---|---|---|---|---|---|
| RNABindPlus | 0.273 | 0.227 | 0.248 | 0.202 | 0.717 |
| SVMnuc | 0.231 | 0.240 | 0.235 | 0.192 | 0.729 |
| COACH-D[*] | 0.221 | 0.252 | 0.235 | 0.195 | 0.663 |
| NucBind[*] | 0.231 | 0.235 | 0.233 | 0.189 | 0.715 |
| aaRNA[*] | 0.484 | 0.166 | 0.237 | 0.214 | 0.771 |
| NucleicNet[*] | 0.371 | 0.201 | 0.261 | 0.216 | 0.788 |
| GraphBind[*a] | 0.303 | 0.171 | 0.218 | 0.168 | 0.718 |
| GraphBind[*b] | 0.463 | **0.294** | **0.358** | **0.322** | **0.854** |
| **CLAPE-RB** | **0.467** | 0.201 | 0.281 | 0.240 | 0.800 |

[*] indicates structure-based models [a] indicates results using predicted protein structures [b] indicates results using experimental protein structures. Results were obtained from Xia et al[47].

Supplementary Table 9: Comparison of ProBert with ESM-v2 as the feature extractor of CLAPE-DB

| Feature extractor | Dataset | AUC | AUPR |
|---|---|---|---|
| ProtBert | TE46 | 0.871 | 0.463 |
| ESM-v2 | TE46 | 0.888 | 0.517 |
| ProtBert | TE129 | 0.881 | 0.411 |
| ESM-v2 | TE129 | 0.908 | 0.528 |

# Figures

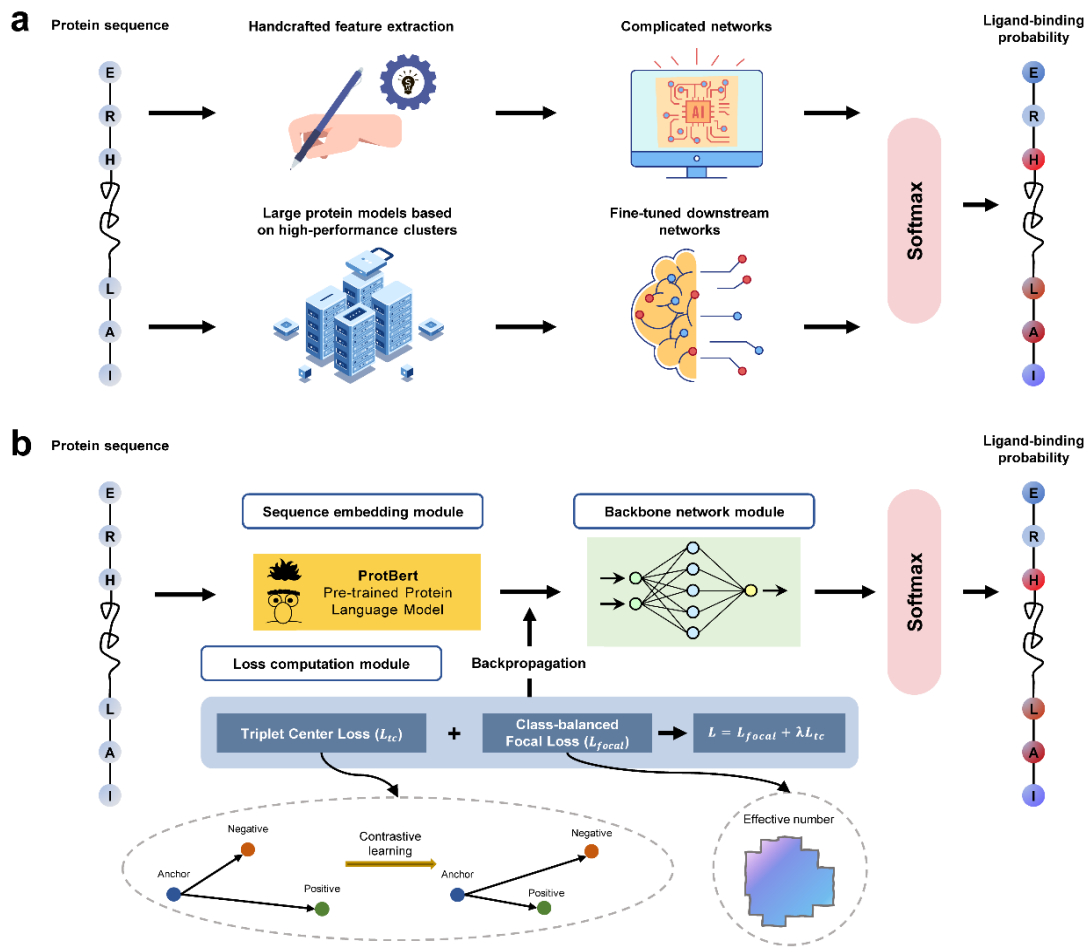

Figure 1: Schematic representations of existing deep learning approaches for DNA-binding site prediction and the overall architecture of the CLAPE model. (a) The upper panels: The conventional classification scheme, which involves a manual feature extraction process followed by a downstream neural network or machine learning model. This approach is time-consuming, as the feature and network design process can take a significant amount of time. The lower panels: The end-to-end classification scheme, which involves a large-scale deep learning model based on a high-performance computational cluster and a downstream fine-tuning model. This approach requires a high cost of computational resources. (b) The CLAPE model consists of three primary modules: the sequence embedding module for generating protein sequence representations using a pre-trained protein language model named ProtBert; the backbone network module for downstream processing of protein embeddings, which is flexible by applying different types of mainstream neural networks, including MLP, CNN, RNN, and GNN; and the loss computation module for computing binary classification focal loss and contrastive loss for backpropagation to update the model parameters of the backbone model. All strategies share a classification module for generating classification scores between 0 and 1, which contains a Softmax function.

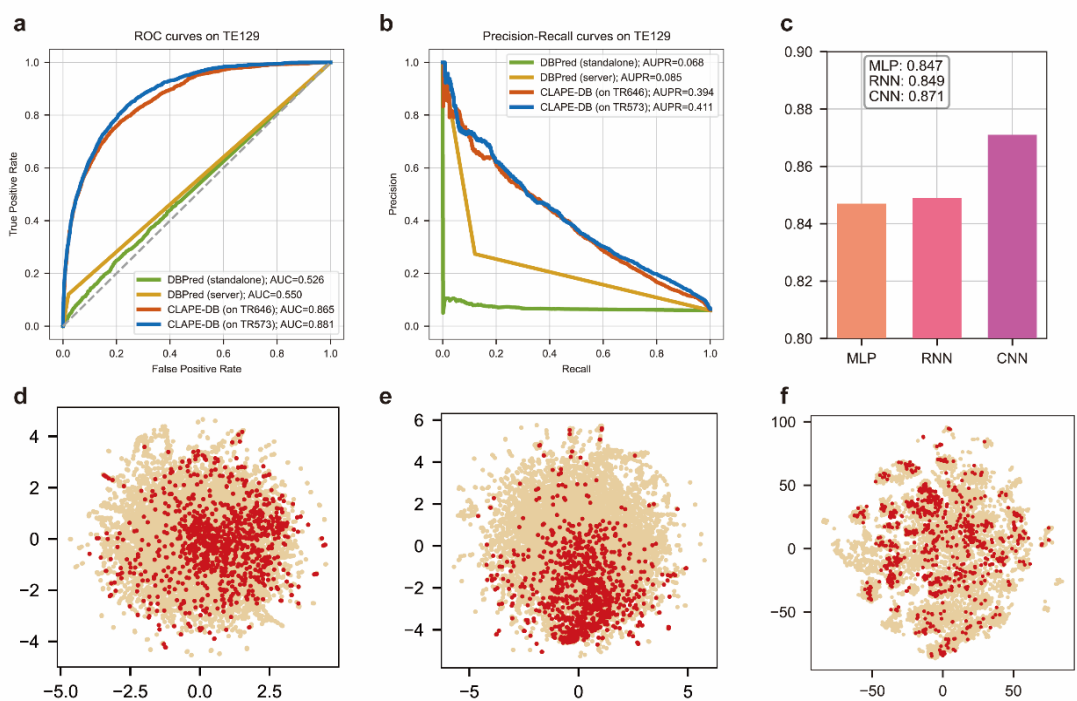

Figure 2: Evaluation of CLAPE-DB model performance. (a) Receiver operating characteristic (ROC) curves of DBPred and CLAPE-DB models. CLAPE-DB showed larger area under ROC curve (AUC) than DBPred, indicating a better generalization ability. (b) Precision-recall (PR) curves of DBPred and CLAPE-DB models. (c) Comparison of different backbone models, where we used an LSTM model to represent RNN. (d) t-SNE dimension reduction plot of the first layer output of a randomly initialized 1DCNN model. (e) t-SNE dimension reduction plot of the first layer output of CLAPE-DB. (f) t-SNE dimension reduction plot of the original sequence features generated by ProtBert. All of (c-f) were tested and plotted using TE46, with cream-colored and red data points indicating non-binding sites and DNA-binding sites, respectively.

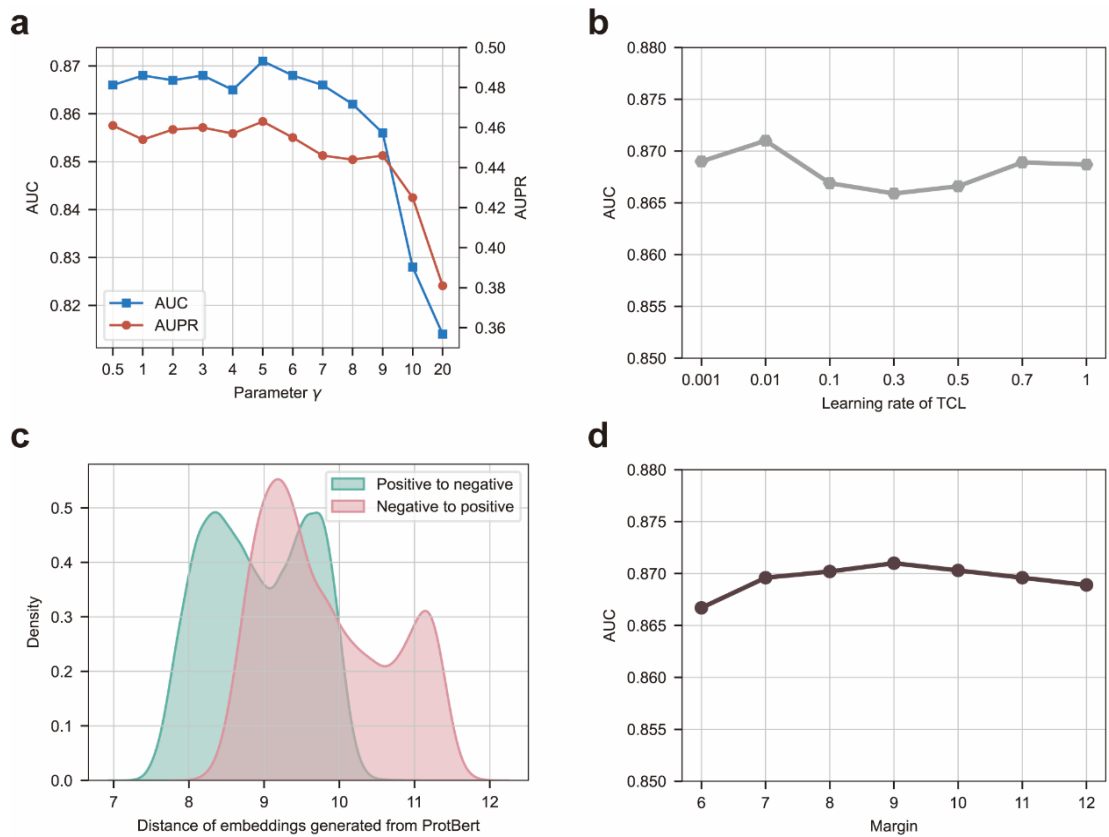

Figure 3: Hyperparameter Optimization of Loss Functions. (a) Trends in the AUC and AUPR metrics with varying parameter $\gamma$. Both metrics reached their maximum values when $\gamma$ was set to 5. (b) Trend in the AUC metric with varying learning rate of the Triplet Center Loss (TCL). The AUC reached its maximum value when the learning rate was set to 0.01. (c) Distance distribution of negative and positive samples, where the distance was defined as the maximum Euclidean distance between a given sample and the sample from the opposite class. The embedding used to calculate the distance was the raw sequence embedding generated from ProtBert. (d) Trend in the AUC metric with varying margin of TCL. The AUC reached its maximum value when the margin was set to 9.

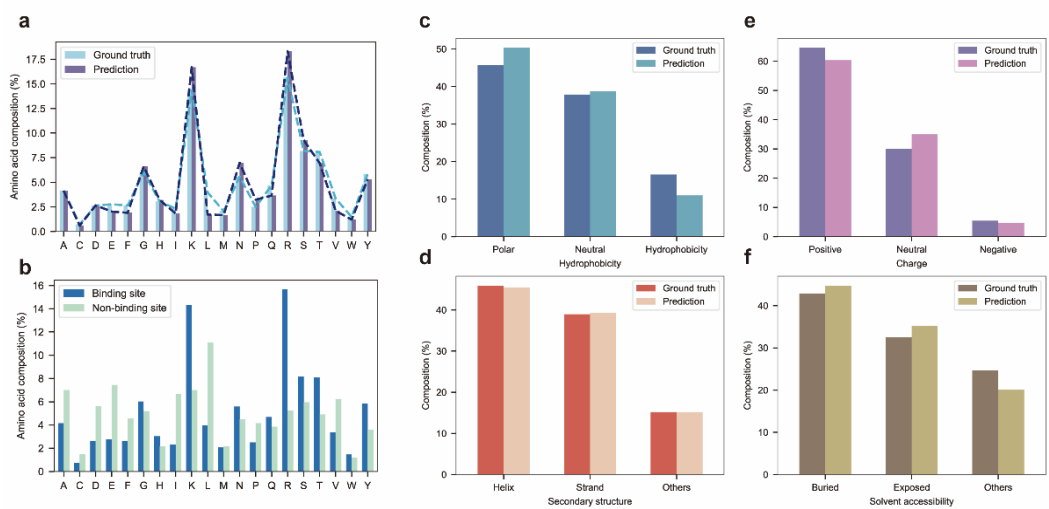

Figure 4: Analysis of amino acid composition and properties. (a) Distribution of amino acid composition in DNA-binding sites and non-binding sites. (b) Comparison of the distribution of experimental DNA-binding sites with predicted binding sites. (c-f) Comparison of the distribution of amino acid physicochemical properties and structural properties of real DNA-binding sites with predicted binding sites. (c-f) represents hydrophobicity, secondary structure, charge, solvent accessibility, respectively.

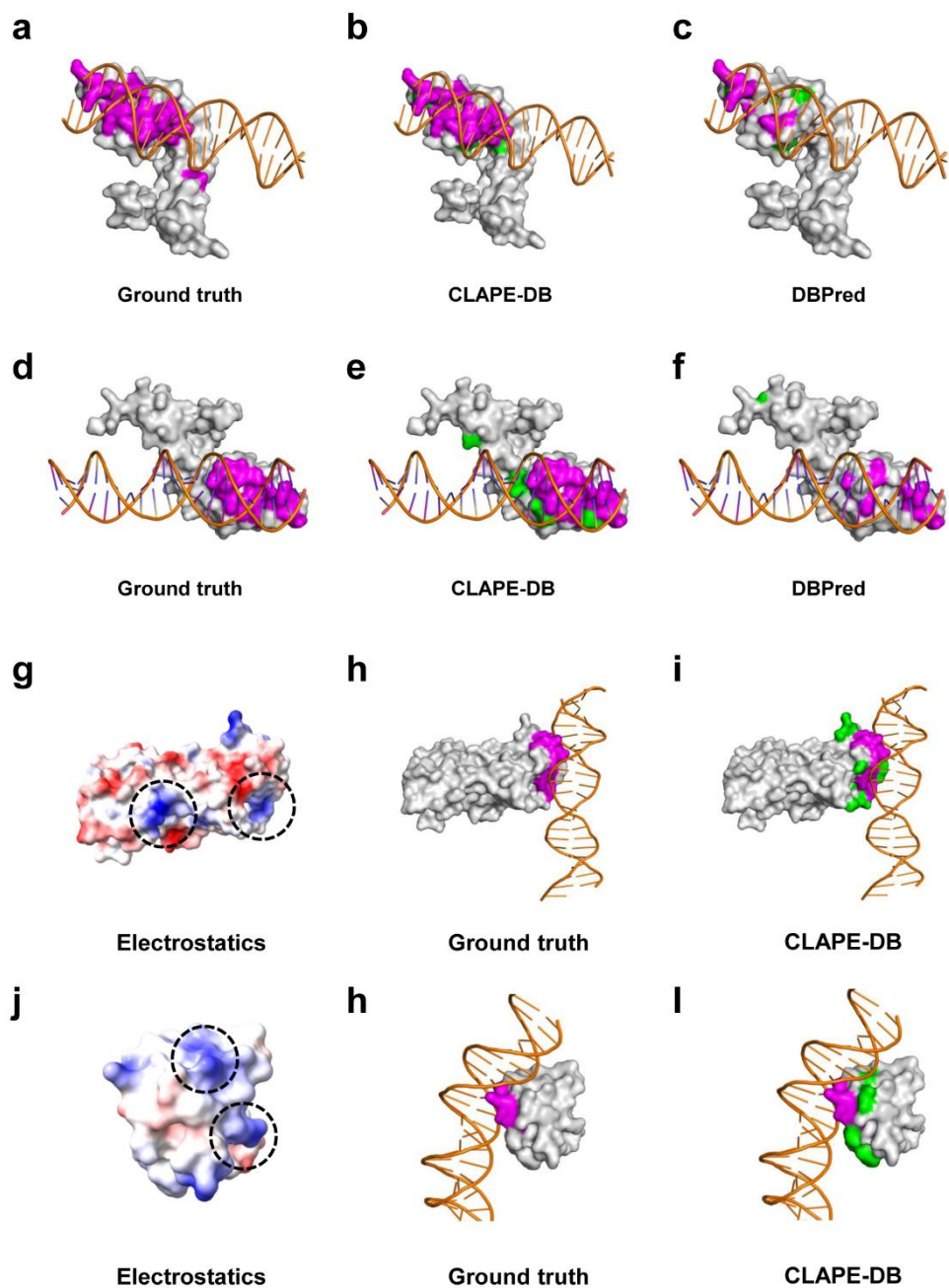

Figure 5: Comparative and empirical case studies. (a-c) Analysis of the DNA-binding sites for protein 5H3R_A, where (a) represents the experimental result, (b) and (c) represent the results predicted by CLAPE-DB and DBPred, respectively. (d-f) Analysis of the DNA-binding sites for protein 6C2S_A, where (d) represents the experimental result, (e) and (f) represent the results predicted by CLAPE-DB and DBPred, respectively. Magenta residues indicate the experimental binding sites and true

positives predicted by the models, green residues indicate the false positives generated by the models, and gray residues indicate non-binding residues. Orange double helix structures represent DNA molecules. (g-h) Comparison of the surface charge distribution and DNA-binding sites for protein 5GPC_A between the experimental and CLAPE-DB predicted results. (j-l) Comparison of the surface charge distribution and DNA-binding sites for protein 5J2Y_A between the experimental and CLAPE-DB predicted results. Blue and red residues indicate positively and negatively charged residues, respectively. The dashed circles in (g) and (i) indicate the human-predicted DNA-binding sites based on electrostatics.

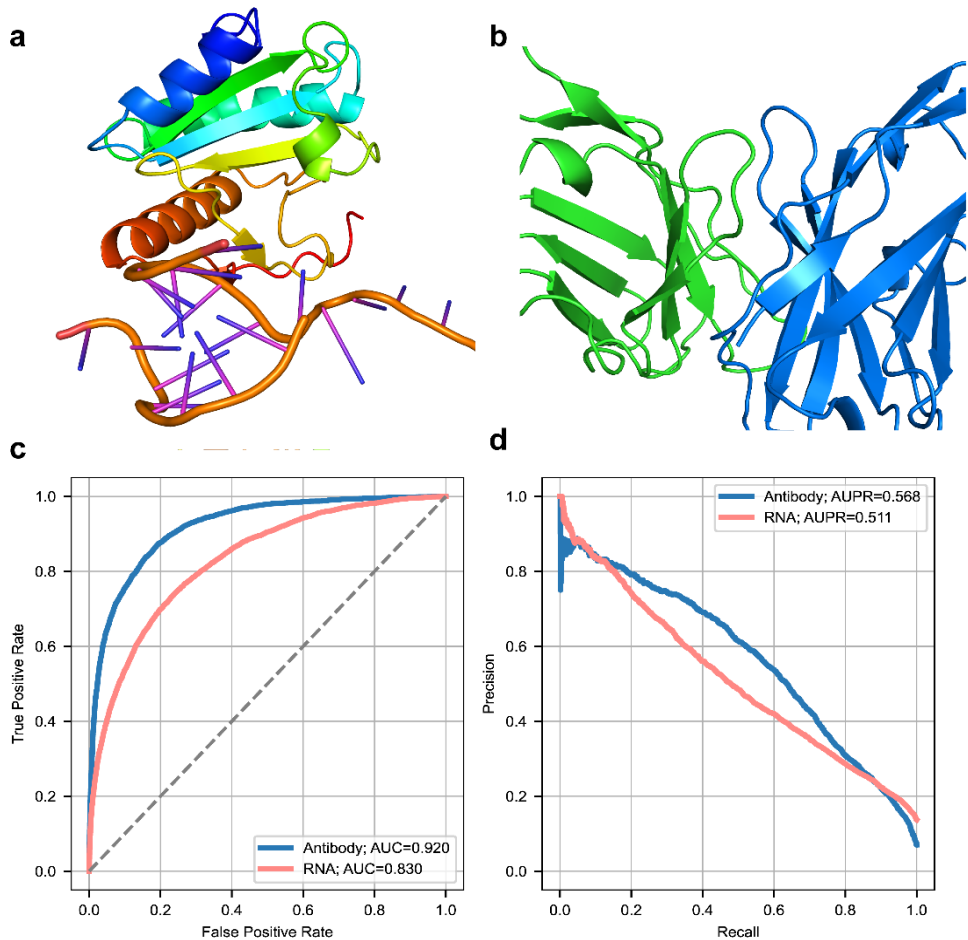

Figure 6: General binding sites prediction ability of CLAPE. (a-b) Binding diagrams of protein-RNA (PDB ID: 5GAN) and antibody-antigen (PDB ID: 1OAY), demonstrating the ability of CLAPE to predict protein-ligand binding sites. (c-d) ROC and PR curves of CLAPE-RB and CLAPE-AB models. CLAPE-RB achieved an AUC of 0.830 and an AUPR of 0.511, while CLAPE-AB achieved an AUC of 0.920 and an AUPR of 0.568.

**Supplementary Figures**

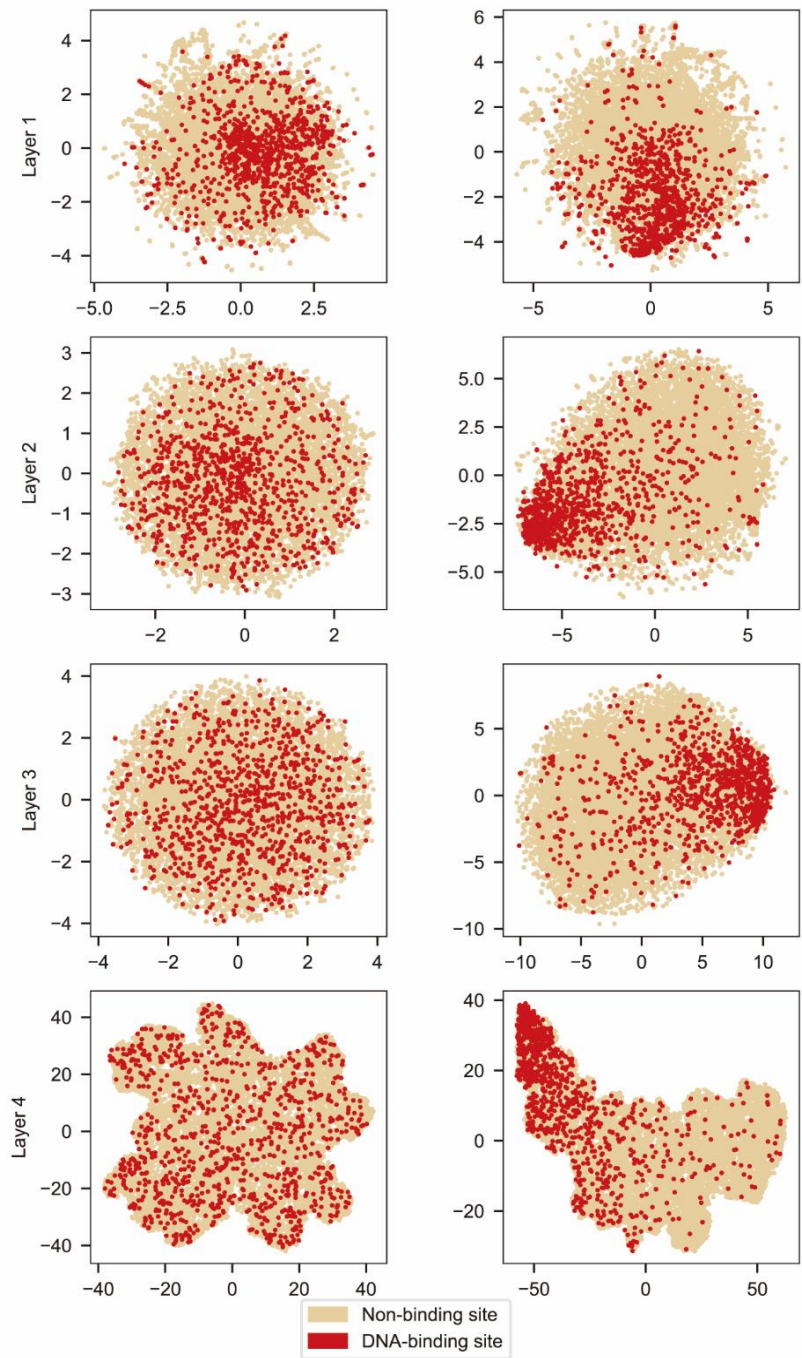

Supplementary Figure 1: t-SNE dimension reduction plot of each convolutional layer of CLAPE-DB, with cream-colored and red data points indicating non-binding sites and DNA-binding sites, respectively.

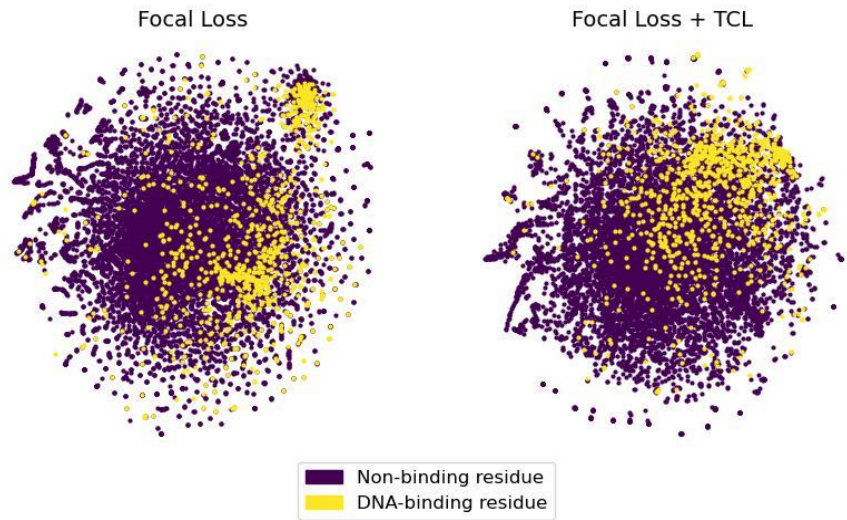

Supplementary Figure 2: t-SNE dimension reduction plot of the output of the first layer using different loss functions.

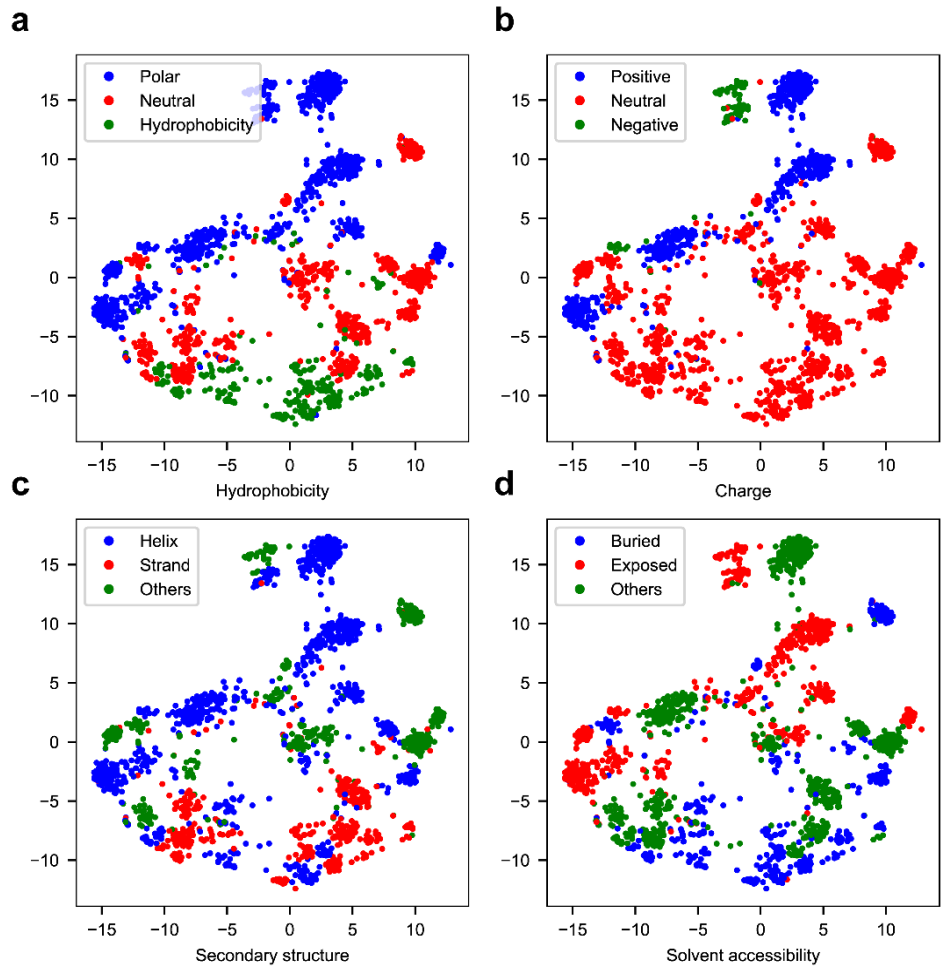

Supplementary Figure 3: t-SNE dimension reduction of raw embeddings of DNA-binding sites concerning properties of different amino acid properties. (a-d) refers to hydrophobicity, charge, secondary structure, and solvent accessibility, respectively.

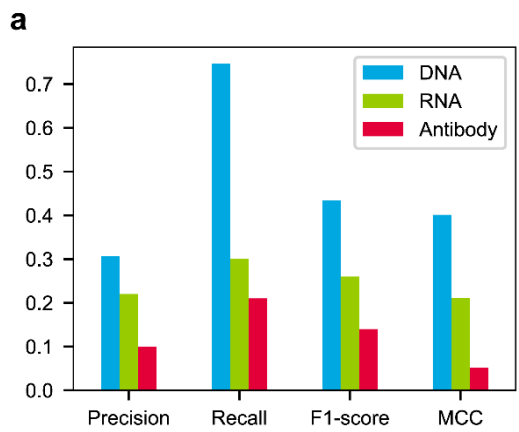 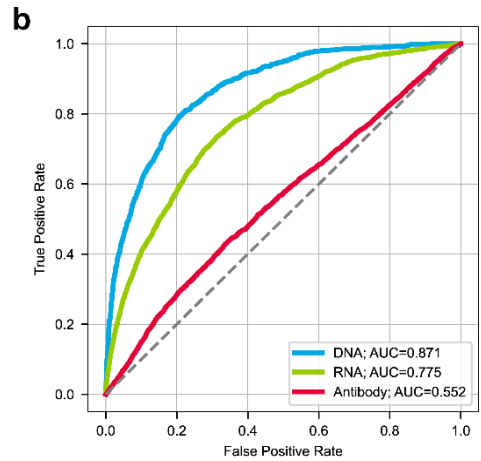

Supplementary Figure 4: The specificity of DNA-binding sites prediction of CLAPE-DB. (a) Prediction performance of different ligand-binding sites of CLAPE-DB. (b) ROC curve of the overall performance of different ligand-binding sites of CLAPE-DB.